\documentclass[times,twocolumn,final,authoryear]{elsarticle}
\usepackage{ycviu}

\usepackage{graphicx}
\usepackage{latexsym}
\usepackage{booktabs}
\usepackage{amsmath,amssymb}
\usepackage[caption=false]{subfig}
\usepackage{array,multirow}
\usepackage{lipsum}
\usepackage{colortbl}
\usepackage{hyperref}

\usepackage{url}
\usepackage{xcolor}

\def\eg{\emph{e.g. }}

\def\ie{\emph{i.e. }}

\definecolor{LightCyan}{rgb}{0.88,1,1}
\DeclareMathOperator{\E}{\mathbb{E}}
\newlength{\dhatheight}

\newcommand{\myparagraph}[1]{\paragraph{#1}}

\usepackage{color,soul}
\soulregister\citet7
\soulregister\citep7

\journal{Computer Vision and Image Understanding}

\title{On the Benefit of Adversarial Training for Monocular Depth Estimation}

\author[1]{Rick \snm{Groenendijk}}
\ead{r.w.groenendijk@uva.nl}
\author[2]{Sezer \snm{Karaoglu}}
\author[1,2]{Theo \snm{Gevers}}
\author[3,1]{Thomas \snm{Mensink}}

\address[1]{University of Amsterdam, Science Park 904, 1098XH Amsterdam, the Netherlands}
\address[2]{3DUniversum, Science Park 400, 1098 XH Amsterdam, the Netherlands}
\address[3]{Google Research, Claude Debussylaan 34, 1082 MD Amsterdam, the Netherlands}

\begin{document}

\begin{frontmatter}

\begin{abstract}
In this paper we address the benefit of adding adversarial training to the task of monocular depth estimation. 
A model can be trained in a self-supervised setting on stereo pairs of images, where depth (disparities) are an intermediate result in a \emph{right-to-left} image reconstruction pipeline. 
For the quality of the image reconstruction and disparity prediction, a combination of different losses is used, including L1 image reconstruction losses and left-right disparity smoothness. 
These are local pixel-wise losses, while depth prediction requires global consistency. 
Therefore, we extend the self-supervised network to become a Generative Adversarial Network (GAN), by including a discriminator which should tell apart reconstructed (fake) images from real images. 
We evaluate Vanilla GANs, LSGANs and Wasserstein GANs in combination with different pixel-wise reconstruction losses.
Based on extensive experimental evaluation, we conclude that adversarial training is beneficial \emph{if and only if} the reconstruction loss is not too constrained. 
Even though adversarial training seems promising because it promotes global consistency, non-adversarial training outperforms (or is on par with) any method trained with a GAN when a constrained reconstruction loss is used in combination with batch normalisation.
Based on the insights of our experimental evaluation we obtain state-of-the art monocular depth estimation results by using batch normalisation and different output scales.
\end{abstract}


\end{frontmatter}

\section{Introduction}\label{intro}
We are interested in estimating depth from single images.
This is fundamentally an ill-posed problem, since a single 2D view of a scene can be explained by many 3D scenes, among others due to scale ambiguity.
Therefore the corresponding 3D scene should be estimated -- \emph{implicitly} -- by looking at the global scene context. 
Using the global context, a model prior can be estimated to reliably retrieve depth from a single image.

To take into account global scene context for single image depth estimation, elaborate image recognition models have been developed~\citep{eigen2014depth, godard2017unsupervised, saxena2006learning}. 
Currently used deep convolutional networks (ConvNets) have enough capacity to understand the global relations between pixels in the image as well as to encode the prior information. 
However, they are only trained with combinations of per-pixelwise losses. 

Generative Adversarial Networks (GANs)~\citep{goodfellow2014generative,isola2017image} force generated images to be realistic and globally coherent. 
They do so by introducing an adaptable loss, in the form of a neural discriminator network that penalises generated output that looks different from real data. 
GANs have seen much research attention in recent years, with ever increasing quality of the generated data (e.g. \citet{karras2019style}). 
Moreover, they have shown success in many tasks, including image reconstruction \citep{isola2017image}, image segmentation~\citep{ghafoorian2018gan}, novel viewpoint estimation~\citep{galama18iclrw} and binocular depth estimation \citep{pilzer2018unsupervised}. 
In this paper, we add an adversarial discriminator network to an existing monocular depth estimation model to include a loss based on global context. 

\begin{figure}[t]
    \centering
	\includegraphics[height=3cm]{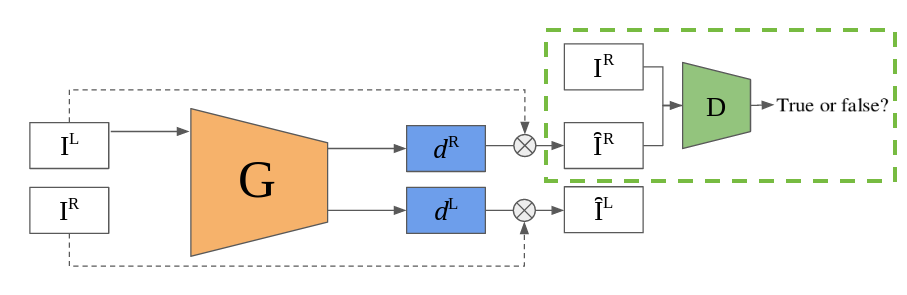}
    \caption{Illustration of the depth prediction architecture. We use the baseline architecture of \citet{godard2017unsupervised}, and extend this by a discriminator (in green) to enable adversarial training.    
    A single input image into the network results in both \emph{left-to-right} and \emph{right-to-left} disparities, which are used to reconstruct both left and right images. 
    The GAN extension, uses the same generator, however a discriminator network D is added to enforce adversarial loss on the generated data of the generator.
    }
    \label{fig:single_gan_architecture}
    \label{fig:base_architecture}
    \label{fig:reconstruction_pipeline}
\end{figure}

Training deep ConvNets requires large datasets with corresponding depth data, preferably dense depth data where ground truth depth is available per pixel. 
These are not always easy to obtain, either due to a complicated acquisition setup or due to sparse depth data of LiDaR sweeps.
To circumvent this, depth prediction has recently been formulated as a novel viewpoint generation task from stereo imagery, where depth is never directly estimated, but just the valuable intermediate in an image reconstruction pipeline. 
\citet{godard2017unsupervised} start from a stereo pair of images, and use the \emph{left} image to estimate the disparities from the \emph{right-to-the-left} image, which is combined with the \emph{right} image to reconstruct the \emph{left} image (see \autoref{fig:reconstruction_pipeline}). Due to the constrained image reconstruction setting, disparities are learned from a single view.
We use this model as our baseline and extend it with a discriminator network, which we train in an adversarial setup.

Previous literature \citep{godard2017unsupervised, yang2018deep, luo2016efficient, mayer2016large} shows impressive performance of depth estimation using mostly engineered photometric and geometric losses. However, most of these loss functions are defined as a sum over per-pixel loss functions. 
The global consistency of the scene context is not taken into account in the loss formulation. 
We address the question: \textit{``To what extent can monocular depth estimation benefit from adversarial training?''}
We do so by studying the influence of adversarial learning on several combinations of photometric and geometric loss functions.

\section{Related Work} 

\myparagraph{Monocular Depth Estimation} Estimating depth from single images is fundamentally different from estimating depth from stereo pairs, because it is no longer possible to triangulate points. 
In a monocular setting, contextual information is required, \eg texture variation, occlusions, and scene context. These cannot reliably be detected from local image patches. 
For example, a patch of blue pixels could either represent distant sky, or a nearby blue coloured object. The global picture thus has to be considered. Global context information can for example be modeled by using manually engineered features~\citep{saxena2009make3d}, or by using CNNs~\citep{eigen2014depth}.

Depth ground truth is expensive and time-consuming to obtain, and the readings might be inaccurate, \eg due to infrared interference or due to surface specularities. 
An alternative is to use self-supervised depth estimation~\citep{garg2016unsupervised,godard2017unsupervised,godard2018digging}, where training data consists of pairs of \emph{left} and \emph{right} images.
A disparity prediction model can be trained, to warp the \emph{left} image into the \emph{right} image, using photometric reconstruction losses. Depth can be recovered from the disparities, by using the camera intrinsics, making depth ground truth data unnecessary at train time.
While stereo pairs are necessary during training, during test time depth can be predicted from a single image. 
In this work we use the work of~\citet{godard2017unsupervised} as a baseline. Their follow-up work is also concerned with depth estimation, yet based on temporal sequences of (monocular) frames~\citep{godard2018digging}. The scope of our paper is on stereoscopic learning of depth.

\myparagraph{GANs for Image Generation}
Depth estimation from single images can be formulated as an image warping or image generation task. Often image generation is done by means of encoder-decoder networks that output newly generated images. Encoder-decoder networks trained using L1 or L2 produce blurry results \citep{pathak2016context, zhao2017loss}, since output pixels are penalised conditioned on their respective input pixels, but never on the joint configuration of the output pixels. 
GANs~\citep{goodfellow2014generative,mirza2014conditional,isola2017image} counteract this by introducing a structured high-level adaptable loss. GANs are used in tasks where generating globally coherent images is important. 
Since monocular depth estimation is largely dependent on how well global contextual information is extracted from the input view, there is reason to believe GANs can be of benefit. 

Pairing the high-level adversarial loss with a low-level reconstruction loss such as L1 may boost performance even more \citep{isola2017image}. This may be due to the fact that the adversarial loss punishes high-level detail, but only slowly updates low-level detail. 
Combining adversarial losses and pixel-based local losses has been shown to work well for a number of tasks, including novel viewpoint estimation~\citep{huang2017beyond,wu2016learning,galama18iclrw}, predicting future frames in a video~\citep{yin2018novel,mathieu2015deep}, and image inpainting~\citep{pathak2016context}.

\myparagraph{GANs for Depth Estimation}
Adversarial losses have already been explored for depth estimation. \citet{chen18arxiv} shows that adversarial training can be beneficial when directly regressing on depth from single images using ground-truth depth data. The authors use a CNN and a CNN-CRF architecture using either L1 or L2 norm as similarity metrics for the predicted depths. Since they only ever use single images during training, they do not exploit scene geometry for more involved geometric losses. 
\citet{kumar18cvprw} predict depth maps from monocular video sequences and successfully use an adversarial network to promote photo-realism between frames. Their generator is composed of two separate networks: A depth network and a pose network. Together these networks enable the authors to generate frames over time. Compared to the current work the problem is less constrained, because the static scene assumption is violated. That is, objects in the scene may themselves move between frames. \citet{pilzer2018unsupervised} suggest to use a cycled architecture for estimating depth, in which two generators and two discriminators jointly learn to estimate depth. Their half-cycle architecture is close to our approach, since it uses a single discriminator. However, the generator requires the input of both left and right images to predict a disparity map and even then does not explicitly enforce consistency between the two images. Concurrently with our research is the work of \citet{aleotti2018generative}, where the method of \citet{godard2017unsupervised} is extended with a vanilla GAN. They address the weighting of loss components and find that a subtle adversarial loss can possibly yield improved performance, albeit marginally. 
In our experiments we find the opposite, none of the used GAN variants improve performance, and we show that small variations of performance could also be explained by initialisation or  attributed to the use of batch normalisation. 
Unlike the methods above the purpose of the current work is to evaluate adversarial approaches when constrained reconstruction losses are used.

We evaluate different GAN objectives for depth estimation, in part based on the results of~\citet{lucic2018gans}, who conclude that no variant is (necessarily) better than others, given a sufficiently large computational budget and extensive hyper-parameter search. We compare the following GAN variants:
\begin{enumerate}
    \itemsep=0mm
	\item Vanilla GAN, with a PatchGAN~\citep{isola2017image} discriminator;
	\item Gradient-Penalty Wasserstein GANs (WGAN-GP)~\citep{arjovsky2017wasserstein,gulrajani2017improved}, which minimises the Wasserstein distance, to overcome the saturating loss of the original GAN formulation;
	\item Least Square GANs (LSGAN)~\citep{mao2017least}, where the sigmoid \emph{real or fake} prediction is replaced by an L2-loss.
\end{enumerate}

While GANs provide a powerful method to output realistic data with an adaptable loss, they are notoriously difficult to train stably~\citep{salimans2016improved}. 
Many strategies to improve training stability of GANs have been proposed, including using feature matching~\citep{ghafoorian2018gan}, using historical averaging~\citep{shrivastava2017learning}, or adding batch normalisation~\citep{ioffe2015batch}.
We include the latter in our GAN variants; initial results have shown that batch normalisation is beneficial for vanilla GANs and LSGANs to counteract internal covariate shift.
\section{Method}
The baseline of our single depth estimation model is the reconstruction-based architecture for depth estimation from \citet{godard2017unsupervised}, which we describe first.
Then, we extend this baseline with an adversarial discriminator network. %

\myparagraph{Depth from Image Reconstruction}
The problem of estimating depth from single images could be formulated as an image reconstruction task, like in \citet{garg2016unsupervised}, where a generator network takes in a single left view image and outputs the left-to-right disparity. 
The right image is reconstructed from the left input image and the predicted disparity. 
As a consequence this network could be trained from rectified left and right image pairs, without requiring ground-truth depth-maps.
At test time depth estimation is based on the disparity predicted from the (single) left image, see~\autoref{fig:base_architecture}.

\citet{godard2017unsupervised} improve on the work of \citet{garg2016unsupervised} by using 
both right-to-left and left-to-right disparities and by using a bilinear sampler \citep{jaderberg2015spatial} to generate images, which makes the objective fully differentiable. 

We first describe in detail the method of \citet{godard2017unsupervised}, our baseline:
The generator \textit{G} uses the left image of the pair to reconstruct both the left and right images. 
Consider a left image $\mathbf{I}^{L}$ and a right image $\mathbf{I}^{R}$. 
Image $\mathbf{I}^{L}$ will be used as the sole input to the generator \textit{G}. 
The generator \textit{G}, however, outputs two disparities $\textit{d}^{L}$ and $\textit{d}^{R}$. 
Using left-to-right disparity $\textit{d}^{R}$ the right image $\hat{\mathbf{I}}^{R}$ can be reconstructed, using warping method $\textit{f}_{w}$:
\begin{equation}
    \hat{\mathbf{I}}^{R} = \textit{f}_{w}(\textit{d}^{R}, \mathbf{I}^{L})
\end{equation}
And similarly, the reconstructed left image $\hat{\mathbf{I}}^{L}$, is obtained by:
\begin{equation}
    \hat{\mathbf{I}}^{L} = \textit{f}_{w}(\textit{d}^{L}, \mathbf{I}^{R})
\end{equation}
A good generator \textit{G} should predict $\textit{d}^{L}$, $\textit{d}^{R}$ such that the reconstructed images ($\hat{\mathbf{I}}^{L}$ and $\hat{\mathbf{I}}^{R}$) are close to the original image pair ($\mathbf{I}^{L}$ and $\mathbf{I}^{L}$). To measure this several image reconstruction losses are used.

\myparagraph{Image Reconstruction Losses}
The quality of the reconstruction is based on multiple loss components, each with different properties for the total optimisation process. 
We combine:
\begin{enumerate}
    \itemsep0em
	\item L1 loss to minimise the absolute per-pixel distance:
	\begin{equation}
	    \mathcal{L}^{l}_{L1} = \frac{1}{\mathrm{N}} \sum_{i, j} \lVert \mathrm{I}_{ij}^{l} - \mathrm{\hat{I}}_{ij}^{l} \rVert.
	\end{equation}
	Note that L1 has been reported to outperform L2~\citep{zhao2017loss}.
	\item Structural similarity (SSIM) reconstruction loss to measure the perceived quality~\citep{wang2004image}:
	\begin{align}
		\mathcal{L}^{l}_{S} &= \frac{1}{\mathrm{N}} \sum_{i, j} \frac{(1 - \mathrm{SSIM}(\mathrm{I}_{ij}^{l}, \mathrm{\hat{I}}_{ij}^{l}))}{2} \\
        \mathrm{SSIM}(x, y) &= \frac{(2\mu_{x}\mu_{y} + c_{1})(2\sigma_{xy} + c_{2})}{(\mu^{2}_{x}\! +
        \! \mu^{2}_{y}\! +\! c_{1})(\sigma^{2}_{x}\! + \! \sigma^{2}_{y} \! + \! c_{2})},
	\end{align}
    where illumination $\mu$ and signal contrast $\sigma$ are computed around centre pixels $x$ and $y$, $c_1= 0.01^{2}$ and $c_2= 0.03^{2}$ overcome divisions by (almost) zeros, these values are set in line with~\citet{godard2017unsupervised, yang2018deep}.

	\item Left-Right Consistency Loss (LR), which enforces the consistency between the predicted \emph{left-to-right} and \emph{right-to-left} disparity maps:
		\begin{equation}
			\mathcal{L}^{l}_{LR} = \frac{1}{\mathrm{N}} \sum_{i, j} \lvert d_{ij}^{l} - d_{i(j + d_{ij}^{l})}^{r} \rvert
		\end{equation}
	\item Disparity Smoothness Loss, which forces smooth disparities, \ie small disparity gradients, unless there is an edge, therefore using an edge-aware L1 smoothness loss:
		\begin{equation}
			\mathcal{L}^{l}_{disp} = \frac{1}{\mathrm{N}} \sum_{i, j} \lvert \partial_{x} d_{ij}^{l} \rvert \ \exp (- \lVert \partial_{x} \mathrm{I}_{ij}^{l} \rVert) + \lvert \partial_{y} d_{ij}^{l} \rvert \ \exp (- \lVert \partial_{y} \mathrm{I}_{ij}^{l} \rVert ),
		\end{equation}
		since the generator outputs disparities at different scales, this loss is normalised by $\tfrac{1}{2^{s}}$ at scale \textit{s} to normalise the output ranges.
\end{enumerate}	

In \citet{yang2018deep} an occlusion loss component was suggested in addition to the other loss terms. Initial experimentation shows no clear benefit of using it for our set-up. Results with the occlusion loss can be found in the supplementary material.
	
The generator network outputs scaled disparities at intermediate layers of the decoder when it is upsampling from the bottleneck layer. 
For each subsequent scale, height and width of the output image is halved.
At each scale the reconstruction loss is computed, and the final reconstruction loss is a combination of the losses at the different scales \textit{s}:
\begin{align}
	\mathcal{L}_{rec} &= \sum_{s=0}^{3} \mathcal{L}_{s} \\ 
	\mathcal{L}_{s} &= \gamma_{L1} \ \mathcal{L}_{L1} + \gamma_{S} \ \mathcal{L}_{S}  + \gamma_{lr} \ \mathcal{L}_{lr} + \gamma_{disp} \ \frac{1}{2^{s}} \mathcal{L}_{disp},
\end{align}
where $\gamma$ weighs the influence of each loss component. For each loss component we use the reconstruction of the \emph{left and right} image, which are defined symmetrically.

\myparagraph{Adversarial Training for Single Image Depth Estimation} \label{chapter:method_sec:adversarial}
We extend the baseline model by a single discriminator network. 
The discriminator network is tasked with discerning between fake and real images on the right side. The schematic is shown in \autoref{fig:single_gan_architecture}, by the green discriminator.

For adversarial training, we combine the reconstruction loss $\mathcal{L}_{rec}$ with the loss functions belonging to specific GAN variants. 
Note that unlike the original formulation of GANs actual data, in the form of the left image $\mathbf{I}^{L}$, is fed to the generator, not noise \textit{z} \citep{mirza2014conditional}. The generator \textit{G} produces two disparities $\textit{d}^{L}$ and $\textit{d}^{R}$. However the discriminator \textit{D} is only presented with the right image $\hat{\mathbf{I}}^{R} = \textit{f}_{w}(\textit{d}^{R}, \mathbf{I}^{L}) = \textit{f}_{w}(G(\mathbf{I}^{L}), \mathbf{I}^{L})$ constructed from $\textit{d}^{R}$. Thus the discriminator \textit{D} examines only (reconstructed) right images $\mathbf{I}^{R}$ and $\hat{\mathbf{I}}^{R}$, to tell apart.
This leads to the following losses:

\newcommand{\ImgL}{\mathbf{I}^{L}}
\newcommand{\ImgR}{\mathbf{I}^{R}}
\newcommand{\hImgR}{\hat{\mathbf{I}}^{R}}

\begin{enumerate}
    \item Vanilla GAN:
    \begin{align}
    \mathcal{L}^{G}_{\textrm{V}} &= - \E \left[ \log D(\hImgR) \right], \\
        \mathcal{L}^{D}_{\textrm{V}} &= \E \left[ \log D(\ImgR) + \log (1 - D(\hImgR)) \right].
    \end{align}
    \item LS-GAN:
    \begin{align}
    \mathcal{L}^{G}_{\textrm{LS}} &= \tfrac{1}{2} \E \left[(D(\hImgR)\!-\!1)^2 \right] \\
        \mathcal{L}^{D}_{\textrm{LS}}  &= \tfrac{1}{2} \E \left[ (D(\mathbf{I}^{R}) - 1)^{2} + D(\hImgR)^{2} \right].
    \end{align}
    \item WGAN:
    \begin{align}
    \mathcal{L}^{G}_{\textrm{W}} &= \E \left[ D(\hImgR) \right] \\
        \mathcal{L}^{C}_{\textrm{W}}  &=  \E \left[ D(\mathbf{I}^{R}) -  D(\hImgR) \right] + \lambda \Omega_{GP}, %
    \end{align}    
    where $\Omega_{GP}$ denotes the gradient penalty with $\lambda$ = 10 from WGAN-GP \citep{gulrajani2017improved}; and where the generator follows the NS-GAN loss~\citep{goodfellow2014generative}.
\end{enumerate}

\noindent The final loss used for training the generator is:
\begin{align}
    \mathcal{L} &= \mathcal{L}_{rec} + \phi_{G} \ \mathcal{L}^{G}_{\textrm{\{V, LS, W\}}},
\end{align}
which combines the reconstruction loss $\mathcal{L}_{rec}$ with the generator part of the GAN loss $\mathcal{L}^{G}$, where $\phi_{G} = 0.1$ weighs its influence.
 
\myparagraph{Evaluation at test time}
At test time only the generator is used to predict the \emph{right-to-left} disparity \textit{d}$^{L}$ at the finest scale, which has the same resolution is the input image.
The predicted disparity is transformed into a depth map by using the known camera intrinsic and extrinsic parameters.

\section{Experiments} \label{section:experimental_evaluation}
In this section we experimentally evaluate the proposed GAN models on two public datasets KITTI and CityScapes.

\begin{table}[t!]
	\centering
	\caption{Evaluation of robustness against initialisation of the networks. Both the baseline model --- our implementation of \citet{godard2017unsupervised}'s model --- and the LSGAN model are trained 10 times. The results indicate that the models are robust against initialisation, albeit some minor variations in the performance remains.}
	\resizebox{0.45\textwidth}{!}{
		\begin{tabular}{@{\extracolsep{4pt}}cl|ccccccc@{}}
			\hline
			& & A & S & R  & Rlog & $\delta^1$ & $\delta^2 $ & $\delta^{3}$      \\ \cline{3-6} \cline{7-9}
						
			& & \multicolumn{4}{c}{lower is better} & \multicolumn{3}{c}{higher is better} \\ \hline
			\multirow{4}{*}{\rotatebox[origin=c]{90}{\parbox[c]{1cm}{\centering Baseline}}} 
			& min & 0.141 & 1.163 & 5.639 & 0.236 & 0.806 & 0.926 & 0.967 \\
			& max & 0.143 & 1.227 & 5.732 & 0.240 & 0.811 & 0.929 & 0.969 \\
			& avg & 0.142 & 1.195 & 5.681 & 0.238 & 0.809 & 0.927 & 0.968 \\
			& std & 0.001 & 0.017 & 0.027 & 0.001 & 0.002 & 0.001 & 0.001 \\\hline
			\multirow{4}{*}{\rotatebox[origin=c]{90}{\parbox[c]{1cm}{\centering LSGAN}}} 
			& min & 0.130 & 1.010 & 5.359 & 0.222 & 0.819 & 0.936 & 0.972 \\
			& max & 0.135 & 1.053 & 5.417 & 0.227 & 0.823 & 0.938 & 0.974 \\
			& avg & 0.133 & 1.038 & 5.388 & 0.225 & 0.821 & 0.937 & 0.973 \\
			& std & 0.001 & 0.014 & 0.019 & 0.001 & 0.001 & 0.001 & 0.001 \\\hline
		\end{tabular}
	}
	
	\label{tab:restarts}
\end{table}

\begin{table}[t]
    \centering
    \caption{Baseline method performance comparison between different generator architecture backbones, VGG and ResNet variants. While the ResNet-50 architecture yields best performance, we use VGG for fair comparison.}
    \resizebox{0.45\textwidth}{!}{
        \begin{tabular}{@{\extracolsep{4pt}}l|c||ccccccc@{}}
        \hline
        \multirow{2}{*}{}&\multirow{2}{*}{\#Params} & A & S & R  & Rlog & $\delta^1$ & $\delta^2 $ & $\delta^{3}$      \\ \cline{3-6} \cline{7-9}
        & & \multicolumn{4}{c}{lower is better} & \multicolumn{3}{c}{higher is better} \\ \hline
        VGG                 & 31.6M     & 0.142 & 1.200 & 5.694 & 0.239 & 0.809 & 0.927 & 0.967 \\
        RN-18            & 20.2M     & 0.146 & 1.260 & 5.771 & 0.243 & 0.801 & 0.924 & 0.967 \\
        \rowcolor{LightCyan}
        RN-50            & 43.9M     & \textbf{0.123} & \textbf{0.936} & \textbf{5.145} & \textbf{0.216} & \textbf{0.843} & \textbf{0.943} & \textbf{0.975} \\
        RN-101           & 62.9M     & 0.124 & 0.971 & 5.280 & 0.219 & 0.840 & 0.942 & 0.974 \\
        \hline
        \end{tabular}
    }
    \label{tab:generator-backbones}
\end{table}

\begin{table*}[t]
    \centering
    \caption{Performance of models using different loss configurations and GAN variants. The best results for each loss configuration are indicated by blue highlighting, the overall best results have been boldfaced. Model configuration 4.a is our implementation of \citet{godard2017unsupervised}.
    We conclude that with the most constraint image reconstruction loss adversarial training does not improve depth estimation, see text for discussion.
    }
    \resizebox{1.0\textwidth}{!}{
        \begin{tabular}{@{\extracolsep{4pt}}lcccc|c|c||ccccccc@{}}
        \hline
        &\multicolumn{4}{c|}{Loss Components} & \multirow{2}{*}{BN} & \multirow{2}{*}{GAN} & ARD & SRD & RMSE  & RMSE log & $\delta < 1.25$ & $\delta < 1.25^{2}$ & $\delta < 1.25^{3}$ \\ 
        \cline{8-11} \cline{12-14}
	&L1 & LR & Disp & SSIM & & & \multicolumn{4}{c}{lower is better} & \multicolumn{3}{c}{higher is better} \\ 
        \hline
        1.a&& & & & \checkmark & Vanilla                        & \cellcolor{LightCyan} 0.810 & 12.442 & 18.245 & 1.999 & \cellcolor{LightCyan} 0.002 & \cellcolor{LightCyan} 0.008 & \cellcolor{LightCyan} 0.020 \\
        1.b&& & & & & LSGAN                              		& 0.893 & 13.826 & 18.816 & 2.468 & 0.000 & 0.000 & 0.000 \\
        1.c&& & & & & WGAN                                      & 0.813 & \cellcolor{LightCyan} 12.310 & \cellcolor{LightCyan} 18.119 & \cellcolor{LightCyan} 1.932 & 0.001 & 0.003 & 0.011 \\
        && & & & & & & & & & & & \\
        2.a&\checkmark & & & & \checkmark &  -                  & 0.200 & 3.149 & 6.795 & 0.289 & 0.760 & 0.904 & 0.956 \\
        2.b&\checkmark & & & & \checkmark & Vanilla             & 0.205 & 3.781 & 7.045 & 0.288 & \cellcolor{LightCyan} 0.771 & \cellcolor{LightCyan} 0.911 & \cellcolor{LightCyan} 0.958 \\
        2.c&\checkmark & & & & \checkmark & LSGAN               & 0.190 & 2.826 & 6.612 & 0.281 & 0.766 & 0.909 & 0.959 \\
        2.d&\checkmark & & & & & WGAN                           & \cellcolor{LightCyan} 0.177 & \cellcolor{LightCyan} 2.398 & \cellcolor{LightCyan} 6.504 & \cellcolor{LightCyan} 0.275 & 0.770 & 0.905 & 0.957 \\
        && & & & & & & & & & & & \\
        3.a&\checkmark & \checkmark & & & \checkmark &     -    & 0.162 & 1.755 & \cellcolor{LightCyan} 5.954 & \cellcolor{LightCyan} 0.253 & 0.789 & 0.922 & \cellcolor{LightCyan} 0.966 \\
        3.b&\checkmark & \checkmark & & & \checkmark & Vanilla  & 0.168 & 2.090 & 6.104 & 0.261 & 0.784 & 0.919 & 0.964 \\
        3.c&\checkmark & \checkmark & & & \checkmark & LSGAN    & \cellcolor{LightCyan} 0.160 & 1.761 & 5.966 & \cellcolor{LightCyan} 0.253 & \cellcolor{LightCyan} 0.792 & \cellcolor{LightCyan} 0.923 & \cellcolor{LightCyan} 0.966 \\
        3.d&\checkmark & \checkmark & & & & WGAN                & 0.170 & \cellcolor{LightCyan} 1.521 & 6.121 & 0.258 & 0.769 & 0.909 & 0.960 \\
        && & & & & & & & & & & & \\
        4.a&\checkmark & \checkmark & \checkmark & \checkmark & &      -                 & 0.142 & 1.200 & 5.694 & 0.239 & 0.809 & 0.927 & 0.967 \\
        4.b&\checkmark & \checkmark & \checkmark & \checkmark & \checkmark & -          & \cellcolor{LightCyan} \textbf{0.132} & \cellcolor{LightCyan} \textbf{1.049} & \cellcolor{LightCyan} \textbf{5.376} & \cellcolor{LightCyan} \textbf{0.224} & \cellcolor{LightCyan} \textbf{0.822} & \cellcolor{LightCyan} \textbf{0.937} & \cellcolor{LightCyan} \textbf{0.974} \\
        4.c&\checkmark & \checkmark & \checkmark & \checkmark & \checkmark & Vanilla    & 0.135 & 1.052 & 5.428 & 0.229 & 0.818 & 0.935 & 0.972 \\
        4.d&\checkmark & \checkmark & \checkmark & \checkmark & \checkmark & LSGAN      & 0.135 & 1.051 & 5.417 & 0.227 & 0.819 & 0.936 & 0.972 \\
        4.e&\checkmark & \checkmark & \checkmark & \checkmark & & WGAN                  & 0.152 & 1.357 & 6.003 & 0.249 & 0.788 & 0.917 & 0.963 \\
                && & & & & & & & & & & & \\\hline
        5&\multicolumn{6}{c||}{\textbf{Training set mean}}                        & 0.361 & 4.826 & 8.102 & 0.377 & 0.638 & 0.804 & 0.894 \\
        \hline
        \end{tabular}
    }
    \label{tab:overview}
\end{table*}

\subsection{Setup} \label{sec:experimental_evaluation-sub:datasets}
\paragraph{Dataset}
For the main set of experiments we use the KITTI~\citep{Geiger2013IJRR} dataset, which contains image pairs of a car driving in various environments, from highways to city centres to rural roads. We follow the Eigen data split~\citep{eigen2014depth} to compare fairly against other methods, which uses 22.6K training images, 888 validation images and 697 test images, which are resized to 256 $\times$ 512. During training no-depth ground truth is used, only the available stereo imagery. For evaluation the provided velodyne laser data is used.

To test if the results generalise to another dataset, we use the CityScapes dataset \citep{cordts2016cityscapes}. This dataset consists of almost 25 thousand images, with 22.9K training images, 500 validation images and 1525 test images. Visual inspection of the CityScapes dataset reveals two things that have consequences for data pre-processing. First, some images contain artifacts at the top and bottom of the images. Second, both left and right cameras capture part of the car on which they are mounted. To compensate the top 50 and bottom 224 rows of pixels are cropped. Cropping is also performed at the sides of the images to retain width and height ratios.

\paragraph{Implementation Details}
The generator-only model from \citet{godard2017unsupervised} is used as a baseline\footnote{Available at \url{https://github.com/rickgroen/depthgan}}.
For all experiments we use an adapted VGG30 generator network architecture, with ${\sim}31.6$M parameters, for fair comparison with other methods~\citep{godard2017unsupervised,pilzer2018unsupervised}.

All models are trained for 50 epochs, in mini-batches of 8, with the Adam optimizer~\citep{kingma2014adam}
The initial learning rate $\lambda_{g} = \lambda_{d}$ is set to $10^{-4}$, and updated using a plateau scheduler~\citep{radford2015unsupervised}.
Data augmentation is done in online fashion, including gamma, brightness, and color shifts and horizontal flipping. In case of the latter left and right images are swapped in order to preserve their relative position.  
Based on recommendations from previous works and some initial experiments, we set $\gamma_{L1}=0.15$,  $\gamma_{\text{S}}=0.85$, $\gamma_{lr}=1.0$, $\gamma_{disp}=0.1$, $\phi_{G}=0.1$.

The discriminators for Vanilla GAN and LSGAN are convolutional network with five layers and for WGAN-GP a 3-layer fully-connected network was used.

\myparagraph{Evaluation}
At test times disparities are warped into depth maps, and the predicted depth is bounded between 0 and 80 metres, which is close to the maximum in the ground truths. Similar to other methods we vertically centre-crop images, see \eg~\citet{garg2016unsupervised}. Ground truth depth data of the Eigen split is sparse.
For quantitative evaluation we use a set of common metrics~\citep{eigen2014depth,godard2017unsupervised, pilzer2018unsupervised, yang2018deep}: Absolute Relative Distance (\textbf{ARD}), Squared Relative Distance (\textbf{SRD}), Root Mean Squared Error (\textbf{RMSE}), log Root Mean Squared Error (\textbf{log RMSE}), and accuracy within threshold \textit{t} (\textbf{$\delta_{t}$}, with $t \in [ 1.25, 1.25^{2}, 1.25^{3} ]$ ). 

Note that the disparity of pixels on the side of the image is ill-defined, the so-called disparity ramps. 
To compensate we use the test image $\mathbf{I}^{L}$ and its flipped version $\mathbf{I}^{L,f}$ to obtain disparity maps $\textit{d}^{L}$ and $\textit{d}^{L,f}$, the latter is flipped again to align with $\textit{d}^L$. 
Both have a disparity ramp, yet on opposite sides of the disparity map.
Therefore we use $\textit{d}^L$, flipped ($\textit{d}^{L,f}$) for the 5\% outer right (left) columns, and average predictions everywhere else.

\myparagraph{Initialisation and Backbone}
In this set of initial experiments we study the robustness of our models to initialisation and the influence of the backbone architecture. 

In this first experiment we study the robustness of our models with respect to the initialisation of the weights and the randomness of the training. 
Therefore, we have trained two of our models 10 times with the same hyper-parameters, namely 
our baseline model, with a VGG backbone, which combines the 4 reconstruction loss components without batch norm and the LSGAN model with batch norm.
The results are shown in~\autoref{tab:restarts}, where we show for each performance measure the minimum, the maximum and the average value, and include the standard deviation.
We observe some small differences in performance, \eg ARD in range $0.141-0.143$ and $\delta<1.25$ in the range $0.806 - 0.811$. 
We conclude that minor differences in performance between models, might in fact be the result of initialisation rather than model design and training choices.

In the second experiment in this section, we study the influence of the backbone network used.
We compare the VGG30 network, used in all other experiments, to variants of a ResNet backbone, using the full reconstruction loss definition (\ie L1, LR, disparity \& SSIM), without any adversarial components.
The results are shown in~\autoref{tab:generator-backbones}. 
We conclude that the ResNet50 architecture yields best performance on the test set and while the shallow ResNet18 is outperformed by the VGG architecture, it is only by relatively little. This is in the line of expectations, since residual learning has been shown to be effective for deep convolutional architectures \citep{he2016deep}.
For the other experiments, however, we use the VGG30 backbone for fair comparison to related work.

\subsection{Loss Components \& GANs}
To address our main research question we have performed an extensive set of experiments using different combinations of loss components paired with three different GAN variants. 
We compare the performance of the model when using different configurations of loss components: L1 Loss, adding Left-Right consistency, adding disparity smoothness, and adding the SSIM loss. For each of the loss component configurations, we also compare the performance of the model without using a GAN, against using a Vanilla GAN, LSGAN or WGAN-GP.
The results are shown in~\autoref{tab:overview}, we refer to experiment indices in parentheses. From the results we observe the following:
\begin{enumerate}
	\item Using only an adversarial loss, without an image reconstruction loss yields imprecise results (1), even worse than using the training set mean (5). The poor performance of the adversarial loss could be explained by the fact that many different disparity maps may reconstruct a correct image. The global coherency loss of the GANs seems to have difficulty converging to locally geometrically viable disparities, we conclude that \emph{both} global and local consistency should be taken into account.
	
	\item All models (with or without GAN) improve when more constraining image reconstruction losses are combined (2, 3, 4); However, where GANs do improve over using just L1 as image reconstruction losses (2.b vs 2.d/2.e), they do not improve when multiple reconstruction losses are used (4.b vs 4.c/d/e).
	
	\item Where WGAN is the best (adversarial) model for the L1 loss, it is the worst (adversarial) model when more constrained losses are used. This could be partly explained by the difficulty of training GAN models, which are highly sensitive to parameters settings and network architectures. 
	
    \item When considering models with a constrained loss (4), adversarial learning boosts performance beyond the baseline (4.a vs 4.c/4.d). However, upon closer inspection, the difference in performance is likely due to the use of batch normalisation (\textit{cf.} 4.b vs 4.c/4.d). Batch normalisation is often used for GANs because it facilitates stable training~\citep{salimans2016improved} and it is the default in many open-source GAN implementations.
    
    \item For the baseline model, we obtain 0.142 ARD (4.a) with our implementation, which is slightly better compared to 0.148 reported by \citet{godard2017unsupervised} (see also~\autoref{tab:other_methods_comparison}). Training with batch normalisation yields an increase of performance to 0.132 (4.b). 
\end{enumerate}

\begin{table}[t]
    \centering
    \caption{The effects of different kinds of normalization on the performance of depth prediction (in ARD) using Vanilla or LSGAN adversarial training. We evaluate no-normalisation (-), instance normalisation (IN), and batch normalisation (BN). We conclude that BN is important for obtaining good results.}
    \resizebox{0.5\textwidth}{!}{
        \begin{tabular}{@{\extracolsep{4pt}}l||ccc|ccc}
        \hline
        \multirow{2}{*}{GAN} & \multicolumn{3}{c}{L1} & \multicolumn{3}{c}{Full} \\
        \cline{2-4} \cline{5-7}
        & - & IN & \multicolumn{1}{c}{BN} & - & IN & BN \\
        \hline
        - & 0.215 & 0.208 & 0.200 & 0.142 & 0.144 & \textbf{0.132} \\
        Vanilla & 0.216 & \textbf{0.183} & 0.205 & 0.143 & 0.145 & 0.135 \\
        LSGAN & 0.216 & 0.184 & 0.190 & 0.143 & 0.142 & 0.135 \\
        \hline
        \end{tabular}
    }
    \label{tab:normalization-table}
\end{table}

\begin{figure*}[p]
	\centering
	\includegraphics[width=\textwidth]{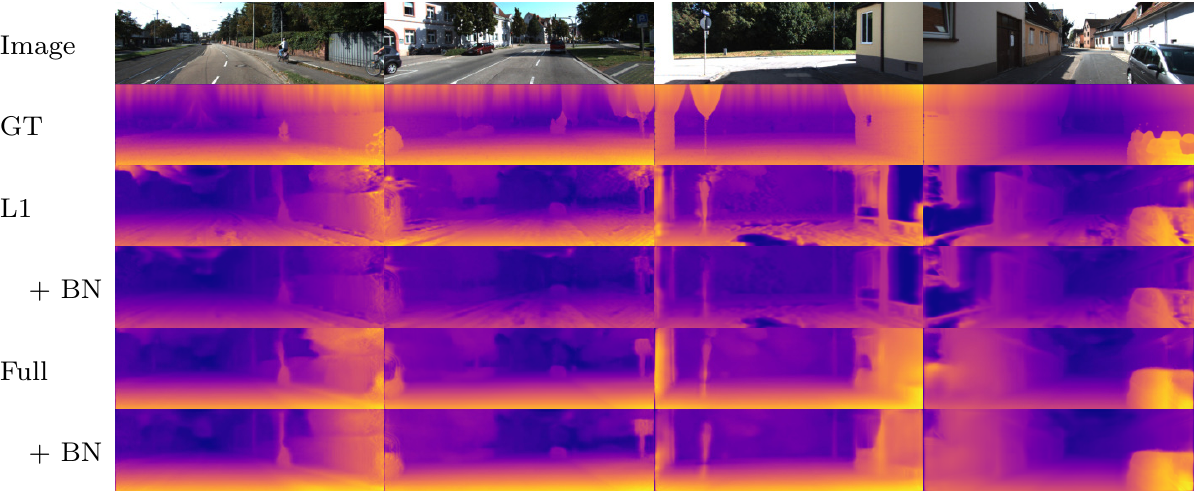}
    \caption{Illustration of the influence of batch normalisation (BN) on depth prediction. 
    Note the inaccurate values in the GT due to in-painting techniques at the top and bottom. 
    BN results in smoother predictions, while keeping small and distinct objects. Best viewed in colour.}
    \label{fig:batch_norm-plot}
\end{figure*}

\myparagraph{Batch \& Instance Normalisation}
In this experiment we evaluate the influence of normalisation strategies on the performance. We compare the L1 and the full reconstruction loss, trained without GAN or with Vanilla / LSGAN. 
The results are shown in ~\autoref{tab:normalization-table}.
From the results, we observe that for any configuration of loss components, performance improves when batch normalisation is used. 
While instance normalisation does not yield better performance for a full loss model, it is beneficial for models trained with a L1 loss. 
We conclude that batch normalisation is important for training for any model. 
Qualitative investigations in~\autoref{fig:batch_norm-plot} show that batch normalisation takes away some granularity in the predicted disparities. 
\begin{figure}[t]
	\centering
	\includegraphics[width=.495\textwidth]{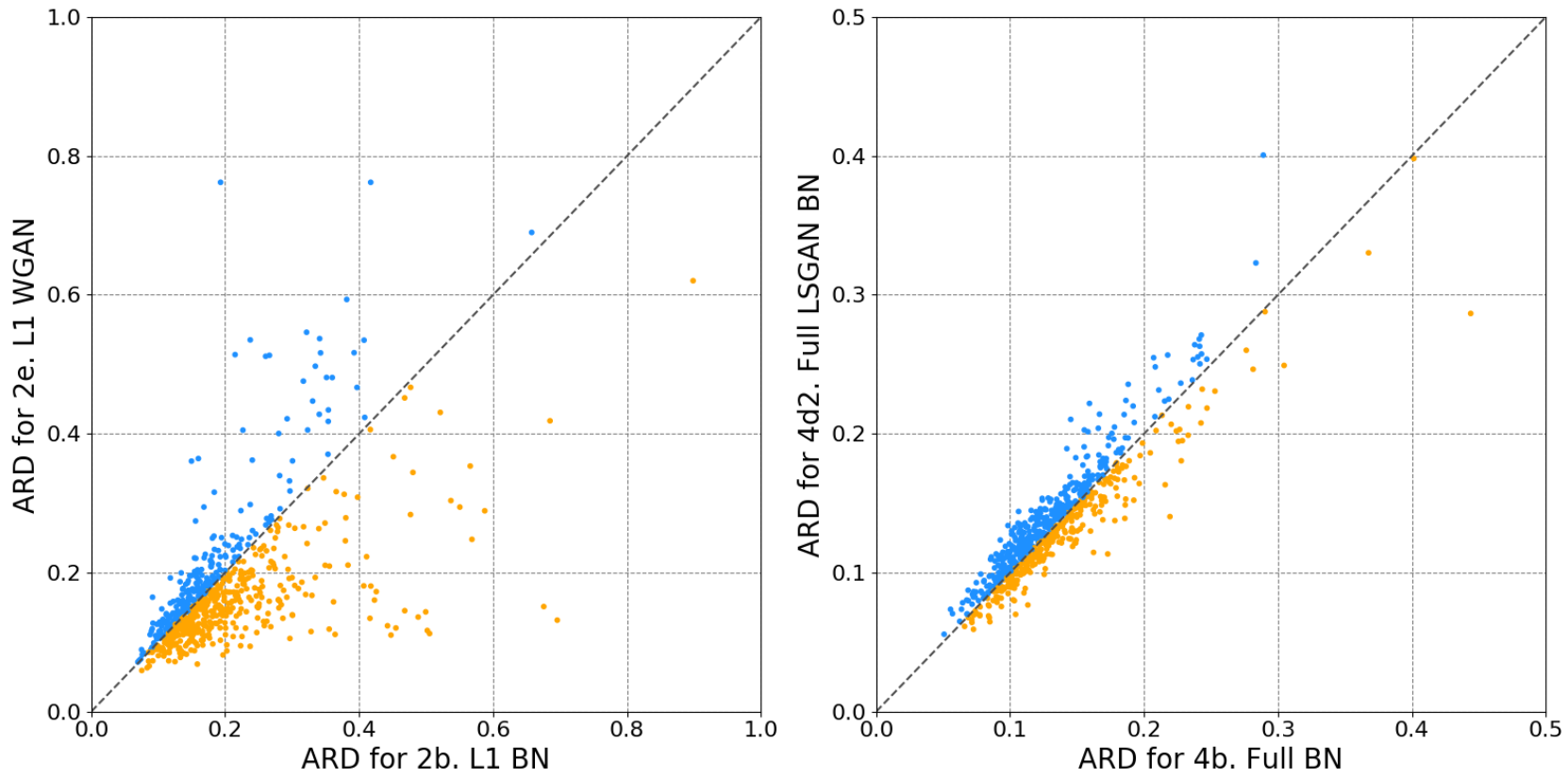}
    \caption{Error scatter plots of the Absolute Relative Distance (ARD, lower is better) for each image, comparing models with and without GAN loss:
    (i) the left plot shows a L1 loss model with BN (entry in~\ref{tab:overview}: 2.b) against a L1 loss WGAN (2.e); 
    (ii) right plot shows a full loss model with BN (4.b) against a full component LSGAN with BN (4.d); 
    The diagonal indicates equal performance for both model. E.g. in (i) the blue dots represent those images for which a L1 loss with BN outperforms a the L1 loss complemented with WGAN. Note the different scaling between plots.
    }
    \label{fig:scatter_plot-figure}
\end{figure}

\newcommand{\hImgL}{$\hat{\mathbf{I}}^{L}$}
\newcommand{\diff}{$(\hat{\mathbf{I}}^{L} - \mathbf{I}^{L})$}
\newcommand{\dL}{$\textit{d}^{L}$}

\begin{figure*}[p]
	\centering
	\includegraphics[width=1\textwidth]{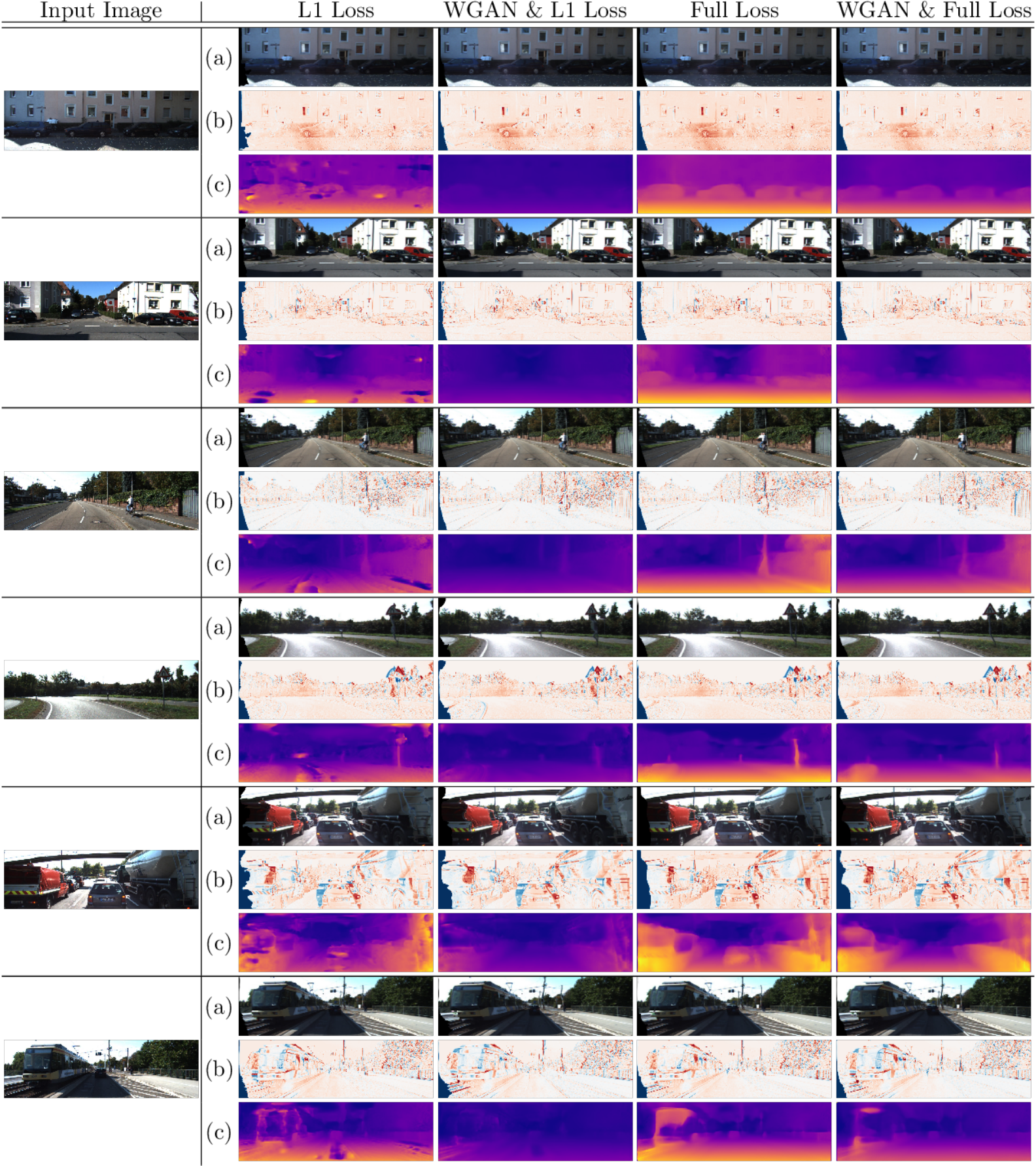}
    \caption{
    Illustrative examples to show the quality of reconstruction and disparity maps for different loss configurations. 
    (a) Reconstructed left images \hImgL; 
    (b) Difference between the ground truth and the reconstructed images \diff, where red indicates regions that are much lighter in the reconstructed image than in the original image, and blue regions indicate spots that are darker. Red and blue areas indicate wrongly reconstructed areas, which yield incorrect disparity values; 
    and
    (c) Generated disparity maps \dL. 
    The top three images are among the top performing images for the full loss model, while the bottom three include failure cases nodes, which achieve poor performance. Also not that images are reconstructed at 256 $\times$ 512 resolution and then upsampled, such that they are less sharp than the input image at full resolution. Best viewed in colour. 
    }
    \label{fig:reconstruction-plot}
\end{figure*}
\begin{figure*}[p]
	\centering
	\includegraphics[width=1\textwidth]{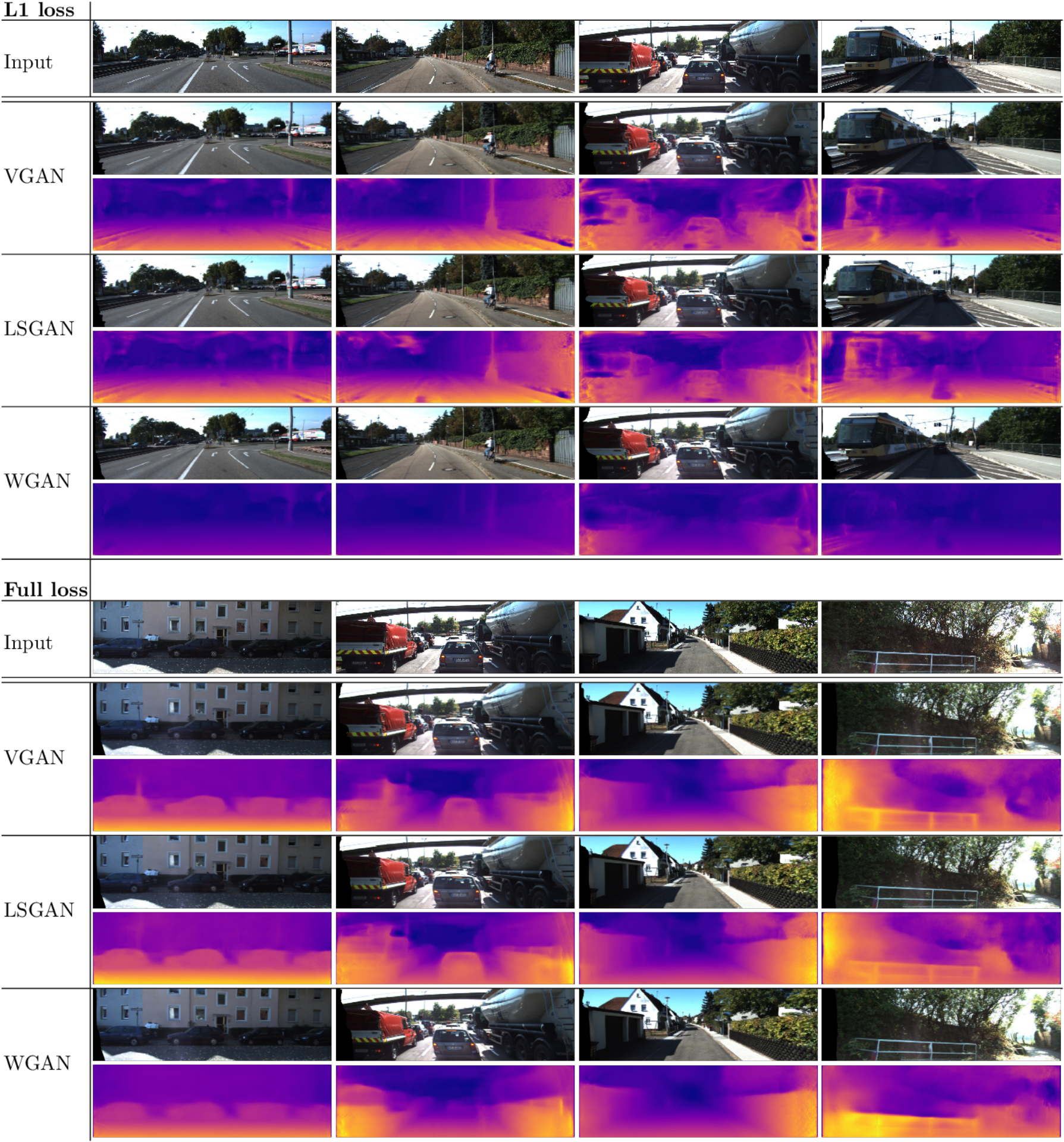}
    \caption{
    Comparison of the three different GAN variants: Vanilla GAN, LSGAN and WGAN, compared for both models trained with only L1 loss (top) and models trained with the full loss (bottom). For each of the image blocks, the top row shows input images $\ImgL$ to the generator. Then for each GAN variant we show the reconstructed left images \hImgL and the right-to-left disparities \dL. Not especially that WGAN severly smooths disparity predictions when paired with only a L1 loss. This is effect is much less pronounced when it is paired with our full loss. Compared with the Vanilla GAN and LSGAN, the WGAN recognises objects in the foreground to a lesser extent (see the second column on the left). Best viewed in colour.
    }
    \label{fig:gan_comparison}
\end{figure*}

\subsection{Reconstructed Image Quality \& GANs}
In the previous section we quantitatively show how performance of models is affected making use of adversarial losses. GANs optimise for photo-realism in reconstructed images, which could lead to well-reconstructed images while the predicted disparities poorly model accurate depth. This is because many disparity maps may reconstruct an image well even though they do not capture depth correctly. 
We evaluate the performance of the non-adversarial model versus GANs at the level of individual images to see if there are cases for which GANs are better suited. 
The results are presented in~\autoref{fig:scatter_plot-figure}. 
When using only the L1 reconstruction loss (\emph{left plane}), the variance in the performance difference between models trained with and without GANs is large (\ie many outliers away from the diagonal).  
However, when using the full reconstruction loss, the scatter is aligned around the diagonal, indicating that both models perform similarly on most images. We have verified the few outliers visually, but do not find a noticeable patterns which indicates better photo-realism for GANs in these cases.

In~\autoref{fig:reconstruction-plot} the reconstructed images and their corresponding disparities are shown for L1, WGAN L1, Full and Full WGAN models. The model trained using only a L1 loss can reconstruct images that have a low pixel-wise loss, but impose no structure on the disparities. This means that there are holes and strong transitions in the predicted disparities. For the WGAN that was trained alongside a L1 loss, it seems the WGANs prefers to predict low disparities. This way images that are very close to the input image are generated. As such, even though the disparities are poor, the reconstructed images are photo-realistic. Adding more geometric loss components that constrain the disparities alleviates these problems.

We also visually compare the performance of different GAN variants in~\autoref{fig:gan_comparison} for models that were trained alongside a L1 loss and a full loss. Again, for models trained with a L1 loss, it can be seen that the WGAN predicts low disparities compared to the other GAN variants. This is beneficial for depth estimation, since it smooths the disparities and prevents under-estimating depth. In a way, the WGAN trained alongside the L1 loss imitates the disparity smoothness loss component. This effect is much less pronounced for models trained with a full loss. 

In a next experiment we compare the image quality (rather than the depth prediction quality) of the reconstructed images, measured by L1, L2, and SSIM scores on the test set.
The results are shown in ~\autoref{tab:image-quality-table}. 
The first row shows the identity mapping, indicating the scores comparing left and right ground truth images directly. In the rest of the table we can see that that are subtle differences between methods in image space. WGAN reconstruction is marginally worse than other methods. The results are interesting, because it seems image reconstruction score has relatively low correlation with depth estimation scores, the image quality of the L1 based models is higher than the full loss, yet the depth prediction is worse.
This implies that there is a need for geometrically founded loss functions, like left-right consistency and disparity smoothness losses to obtain better depth predictions.

\begin{table}[t]
    \centering
    \caption{Comparison of image reconstruction quality measured in L1, L2 norms and SSIM. The identity row indicates the losses that would be observed when comparing left and right images. This would happen when the generator learns to predict only zero-valued disparity maps, such that images are not warped using the disparities.}
    \resizebox{.5\textwidth}{!}{
        \begin{tabular}{@{\extracolsep{4pt}}l||ccc}
        \hline
        \multirow{2}{*}{Method} & \multicolumn{1}{c}{L1} & \multicolumn{1}{c}{L2} & \multicolumn{1}{c}{SSIM} \\
        \cline{2-3} \cline{4-4}
        & \multicolumn{2}{c}{lower is better} & \multicolumn{1}{c}{higher is better} \\
        \hline \hline
        Identity                    & 0.1102 & 0.0387 & 0.4757 \\
        \hline
        L1 loss - base              & 0.0499 & 0.0102 & 0.7110 \\
        L1 loss - Vanilla GAN       & \bf 0.0495 & \bf 0.0101 & \bf 0.7130 \\
        L1 loss - LSGAN             & 0.0521 & 0.0112 & 0.7003 \\
        L1 loss - WGAN              & 0.0525 & 0.0111 & 0.6945 \\
        \hline
        Full loss - base            & \bf 0.0514 & \bf 0.0110 & \bf 0.7183 \\
        Full loss - Vanilla GAN     & 0.0521 & 0.0112 & 0.7115 \\
        Full loss - LSGAN           & 0.0518 & 0.0111 & 0.7125 \\
        Full loss - WGAN            & 0.0550 & 0.0126 & 0.6936 \\
        \hline
        \end{tabular}
    }
    \label{tab:image-quality-table}
\end{table}

\begin{table}[t]
    \centering
    \caption{Performance of depth estimation using different numbers of scales during training, measured in ARD.
    We conclude that using just 1 or 2 scales suffice for good performance.}
    \resizebox{.5\textwidth}{!}{
        \begin{tabular}{@{\extracolsep{4pt}}cccc|c|c||ccc}
        \hline
        \multicolumn{4}{c|}{Loss Components} & \multirow{2}{*}{BN} & \multirow{2}{*}{GAN} & \multicolumn{3}{c}{Scales}\\
         L1 & LR & Disp & SSIM & & & 4 & 2 & 1 \\ \hline
        \checkmark & \checkmark & & &  &                             			& 0.191 & 0.190 & 0.187 \\
        \checkmark & \checkmark & & &  \checkmark &                  		& 0.162 & 0.145 & \bf 0.138 \\
        \checkmark & \checkmark & & &  \checkmark & Vanilla          		& 0.168 & 0.156 & \bf 0.138 \\
        \checkmark & \checkmark & & &  & WGAN                        			& 0.170 & 0.154 & 0.181 \\
        \hline
        \checkmark & \checkmark & \checkmark & \checkmark &  &                       & 0.142 & 0.137  & 0.137 \\
        \checkmark & \checkmark & \checkmark & \checkmark & \checkmark &            & \emph{0.132} & \bf 0.128 & \emph{0.131} \\
        \checkmark & \checkmark & \checkmark & \checkmark &  \checkmark & Vanilla    & 0.135 &  0.136 & \emph{0.132} \\
        \checkmark & \checkmark & \checkmark & \checkmark &  & WGAN                  & 0.152 & 0.152 & 0.154 \\
        \hline
        \end{tabular}
    }
    \label{tab:output-scales}
\end{table}

\subsection{Output scales}
In this set of experiments, we vary the number of output disparity predictions from 4 (as used in~\citet{godard2017unsupervised}) to 1 (as used in~\citet{pilzer2018unsupervised}).
We do so for two settings of the reconstruction loss, using only L1 + LR loss, similar to~\citet{pilzer2018unsupervised} and using the full reconstruction loss, similar to~\citet{godard2017unsupervised}.
The results are shown in~\autoref{tab:output-scales}, using the ARD performance measure.
From the results we observe that reducing disparity output scales contributes positively to depth estimation quality, this holds especially in the case of using just the L1 + LR loss.
The intuition is that forcing the network to output coherent disparities early on can over-regularise disparities: With 4 output scales the  smallest resolution of disparities is just 32 $\times$ 64, which acts as a (too) strict regulariser. 

\begin{table*}[p]
    \centering
    \caption{Evaluation of KITTI trained models on the CityScapes dataset~\citep{cordts2016cityscapes}.}
    \resizebox{1.0\textwidth}{!}{
        \begin{tabular}{@{\extracolsep{4pt}}lcccc|c|c|c||ccccccc@{}}
        \hline
        &\multicolumn{4}{c|}{Loss Components} & \multirow{2}{*}{S} & \multirow{2}{*}{BN} & \multirow{2}{*}{GAN} & ARD & SRD & RMSE  & RMSE log & $\delta < 1.25$ & $\delta < 1.25^{2}$ & $\delta < 1.25^{3}$ \\ 
        \cline{9-12} \cline{13-15}
	    &L1 & LR & Disp & SSIM & & & & \multicolumn{4}{c}{lower is better} & \multicolumn{3}{c}{higher is better} \\ 
        \hline
        1.a & \checkmark &  & & & 4 & &                              & 0.409 & 11.281 & 20.515 & 0.569 & 0.310 & 0.605 & 0.781 \\
        1.b & \checkmark &  & & & 4 & & WGAN                         & 0.329 & 7.557 & 19.646 & 0.512 & 0.370 & 0.644 & 0.803 \\
        & & & & & & & & & & & & \\
        \rowcolor{LightCyan}
        2.a&\checkmark & \checkmark & \checkmark & \checkmark & 1 & &                   & \textbf{0.308} & \textbf{6.133} & \textbf{18.245} & \textbf{0.472} & \textbf{0.335} & \textbf{0.664} & \textbf{0.834} \\
        2.b&\checkmark & \checkmark & \checkmark & \checkmark & 1 & & WGAN              & 0.321 & 6.687 & 18.791 & 0.491 & 0.339 & 0.657 & 0.822 \\ \hline
        3.a&\checkmark & \checkmark & \checkmark & \checkmark & 4 & &                       & 0.324 & 6.927 & 19.309 & 0.505 & 0.323 & 0.642 & 0.808 \\
        3.b&\checkmark & \checkmark & \checkmark & \checkmark & 4 & & LSGAN                 & 0.324 & 6.889 & 19.060 & 0.501 & 0.322 & 0.644 & 0.810 \\
        3.c&\checkmark & \checkmark & \checkmark & \checkmark & 4 & \checkmark & LSGAN      & 0.310 & 6.310 & 18.576 & 0.479 & 0.337 & 0.660 & 0.826 \\
        \hline
        \end{tabular}
    }
    \label{tab:cityscapes_generalizability}
\end{table*}

\begin{table*}[p]
    \centering
    \caption{Comparison with state-of-the-art methods, both fully supervised~\citep{yang2018deep} and using \emph{left-right correspondence} as supervision~\citep{pilzer2018unsupervised,godard2017unsupervised}. The improved work of \citet{godard2018digging} is also shown, which reports better performance of their version 1 (i.e. \citet{godard2017unsupervised}) baseline. The improved version 1 has a ResNet50 backbone and up-samples low-resolution disparity maps before evaluating the loss. We only consider models that have not been pre-trained. 
    Our model outperforms the work of~\citet{godard2017unsupervised}, due to using batch normalisation and better use of disparity prediction scales.}
    \resizebox{\textwidth}{!}{
        \begin{tabular}{@{\extracolsep{4pt}}lcc||ccccccc@{}}
        \hline
        \multirow{2}{*}{Method}  & \multirow{2}{*}{} & \multirow{2}{*}{Trained on}
        & ARD & SRD & RMSE & RMSE log & $\delta < 1.25$ & $\delta < 1.25^{2}$ & $\delta < 1.25^{3}$ \\ 
        \cline{4-7} \cline{8-10}
        & & & \multicolumn{4}{c}{lower is better} & \multicolumn{3}{c}{higher is better} \\
        \hline
        \multicolumn{3}{l||}{\bf Supervised using LiDAR depth} & & \\     
        \citet{saxena2006learning}& LiDAR & K               & 0.280 & - & 8.734 & - & 0.601 & 0.820 & 0.926 \\
        \citet{eigen2014depth} & LiDAR & K                  & 0.203 & 1.548 & 6.307 & 0.282 & 0.702 & 0.890 & 0.958 \\        
        \citet{yang2018deep} & LiDAR & K                    & \textbf{0.097} & \textbf{0.734} & \textbf{4.442} & \textbf{0.187} & \textbf{0.888} & \textbf{0.958} & \textbf{0.980} \\[3mm]
        \multicolumn{3}{l||}{\bf Supervised using Left-Right Correspondence} & & \\                   
        \citet{pilzer2018unsupervised} &  & K                   & 0.152 & 1.388 & 6.016 & 0.247 & 0.789 & 0.918 & 0.965 \\
        \citet{godard2017unsupervised} & VGG & K                & 0.148 & 1.344 & 5.927 & 0.247 & 0.803 & 0.922 & 0.964 \\
        \citet{godard2018digging} & R50, video based, v1 & K    & 0.133 & 1.142 & 5.533 & 0.230 & 0.830 & 0.936 & 0.970 \\
        \citet{godard2018digging} & R18, v2 & K                 & \bf 0.130 & \bf 1.144 & \bf 5.485 & \bf 0.232 & \bf 0.831 & \bf 0.932 & \bf 0.968 \\[3mm]
        \multicolumn{3}{l||}{\bf This paper} & & \\                   
        \bf Baseline & VGG, S4 & K                              & 0.142 & 1.200 & 5.694 & 0.239 & 0.809 & 0.927 & 0.967 \\
        \bf Optimised settings & VGG, BN + S2 & K               & 0.128 & 1.026 & 5.313 & 0.222 & 0.830 & 0.939 & 0.973 \\
        \bf ResNet - Baseline & R50, S4 & K                     & 0.123 & 0.936 & 5.145 & 0.216 & 0.843 & 0.943 & \bf 0.975 \\
        \bf ResNet - Optimised settings & R50, BN + S2 & K                         & \bf 0.122 & \bf 0.928 & \bf 5.119 & \bf 0.215 & \bf 0.847 & \bf 0.945 & \bf 0.975 \\
        \hline
        \end{tabular}
    }
    \label{tab:other_methods_comparison}
\end{table*}

\subsection{Generalising from KITTI to CityScapes} \label{sec:experimental-evaluation_sub:general}
In this set of experiments we evaluate our models on the CityScapes dataset, after training on the KITTI dataset.
The goal is to see if the insights and results generalise to this novel domain. 
Results are depicted in~\autoref{tab:cityscapes_generalizability}. 
The results indicate similar behaviour on the CityScapes dataset as on the KITTI dataset: when the reconstruction losses are sufficiently constrained, adversarial training does not improve the performance. We obtain the best generalising results by using a combination of all four image reconstruction losses, trained at a single scale, using batch normalisation.

\myparagraph{Comparison to State-of-the-Art}
In the final set of experiments, we compare the performance of our current work to a few state-of-the-art methods on KITTI, see~\autoref{tab:other_methods_comparison}. For comparison, we report performance on the Eigen test set, using centre cropping~\citep{garg2016unsupervised}.
For reference we include a few seminal works on monocular depth estimation, when using the LiDAR data during training~\citep{saxena2006learning,eigen2014depth} and one of the newest methods~\citep{yang2018deep}. Then we compare our performance to the work of \citet{godard2017unsupervised}, \citet{godard2018digging}, and \citet{pilzer2018unsupervised}, since these serve as baselines and inspiration for the current work.
While adding adversarial losses to take into account scene context does not improve over a combination of reconstruction based lasses, using batch normalisation and just 2 output scales significantly boost performance.

\section{Conclusions}
This work has sought to investigate whether using adversarial losses benefits the estimation of depth maps in a monocular setting. 
For many tasks where global consistency is important, adversarial training improves image reconstruction tasks.
However, after extensive experimental evaluation, we conclude that adversarial training is beneficial in monocular depth estimation \emph{if and only if} the reconstruction loss does not impose too many constraints on reconstructed images. When more involved geometrically or structurally inspired losses are introduced, adversarial training contributes hardly to the quality of the predicted depth maps and may even be harmful.

Based on our extensive experiments we also conclude that:
\begin{itemize}
    \item[(i)] Batch normalisation improves depth prediction quality significantly; and
    \item[(ii)] evaluating reconstruction losses at many output scales over-regularises the disparity at the final scale, this effect is stronger when the loss function itself is more constrained by SSIM and left-right consistency components;
\end{itemize}
Using those two insights, we have been able to set a new state-of-the-art monocular depth prediction based on reconstruction losses, improving from 0.148 ~\citep{godard2017unsupervised} to 0.128 (using batch norm and 2 output scales).

Future research could investigate the influence of specific architectures of the discriminator network and the use of conditional GANs for guiding monocular depth estimation.

\section*{Acknowledgements}
This research was supported in part by the Dutch Organisation for Scientific Research via the VENI grant~\textbf{What \& Where} awarded to Dr. Mensink.

\bibliographystyle{model2-names}
\bibliography{refs}

\end{document}